\documentclass[aps,twocolumn,pra,superscriptaddress,nofootinbib]{revtex4-2}

\usepackage{amsmath}

\usepackage{amssymb}

\usepackage{booktabs}
\usepackage{longtable}
\usepackage{array}

\newcolumntype{P}[1]{>{\raggedright\arraybackslash}p{#1}}

\usepackage{mathrsfs}
\usepackage{bm, dsfont}
\usepackage[usenames,dvipsnames]{color}
\usepackage{colortbl}
\usepackage[table]{xcolor}
\usepackage{enumitem}
\usepackage{units}
\usepackage{graphicx}
\usepackage{natbib}
\usepackage[colorlinks=true,linkcolor=blue,citecolor=blue,urlcolor=blue]{hyperref}
\usepackage{cleveref}
\usepackage{hypcap}
\usepackage{verbatim, float}
\usepackage{psfrag}
\usepackage{tikz}
\usetikzlibrary{arrows.meta, automata,
                positioning,
                quotes}
\usetikzlibrary{arrows.meta, shapes.geometric,calc}
\usepackage[normalem]{ulem}
\usepackage{physics}		
\usepackage{amsmath}
\usepackage{amsthm}

\newtheorem{lemma}{Lemma}

\newcommand{\id}{\mathds{1}}

\newcommand{\dA}{\mathds{A}}

\newcommand{\cE}{\mathcal{E}}

\begin{document}
\author{Pavel Sekatski}
\affiliation{Department of Applied Physics, University of Geneva, Switzerland}
\author{Jef Pauwels}
\affiliation{Department of Applied Physics, University of Geneva, Switzerland}
\affiliation{Constructor University, 28759 Bremen, Germany}

\title{All Entangled States are Nonlocal and Self-Testable in the Broadcast Scenario}

\begin{abstract}
Entanglement and Bell nonlocality are known to be inequivalent: there exist entangled states that admit a local hidden-variable model for all local measurements. Here we show that this gap disappears in a minimal broadcast extension of the Bell scenario. Assuming only the validity of quantum theory, we prove that for every entangled state  $\rho_{AB}$ there exist local broadcasting maps and local measurements such that the resulting four--partite correlations cannot be reproduced by any broadcast network whose source is separable across the $A|B$ cut. Thus, all entangled states are broadcast nonlocal in quantum theory. In addition, we show that all (also mixed) multipartite states can be broadcast-self-tested, according to a natural operational definition.
\end{abstract}

\maketitle

\textit{Introduction.---} It is by now well understood that entanglement is not always sufficient for Bell nonlocality. Werner showed that there exist bipartite mixed (Werner) states $\rho_{AB}$ that are entangled but admit a local hidden variable model that reproduces the statistics of all local projective measurements~\cite{Werner89}, which was later extended to all measurements~\cite{Barrett2002,Renner2023,Zhang2023}. 
In other words, there are entangled states for which every Bell experiment looks classical. This reveals a fundamental gap between the intrinsic resource content of entangled systems and its empirical consequences -- correlations observable in a Bell test. Unlike the latter, entanglement is a theoretical feature of the underlying quantum description not directly empirically accessible.

A natural question is therefore whether one can slightly modify the original Bell scenario so that every entangled state can be witnessed through correlations. Several answers are known once one is willing to introduce trusted quantum devices. For instance, in the ``textbook" entanglement detection scheme, Alice and Bob are given tomographically complete fully trusted measurements and they can then reconstruct the shared state and certify entanglement \cite{Doherty04,Guhne2009}.  Alternatively, Buscemi showed that all entangled states can be detected if Alice and Bob are provided with the appropriate set of trusted state-preparation devices~\cite{Buscemi12}, which was later formalized~\cite{Branciard2012} into the framework of measurement–device–independent entanglement witnesses (MDI–EWs). In these, and similar approaches \cite{Masanes08}, entanglement detection ultimately relies on theoretical modeling of some auxiliary quantum devices (state sources or measurements), and hence depart from a fully device–independent or strictly empirical viewpoint.

\begin{figure*}[t]
\centering
\includegraphics[width=0.9\textwidth]{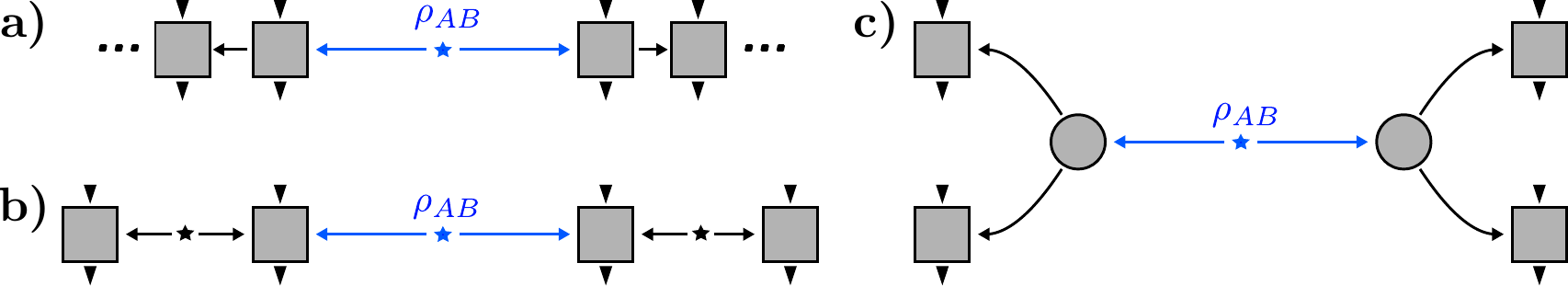}
\caption{Modifications of the standard bipartite single-round Bell scenario, aiming at device-independent entanglement detection of the source $\rho_{AB}$ (blue). The auxiliary devices (black) are untrusted, but assumed to follow quantum mechanics. 
\textbf{(a)} Sequential-measurement scenario: each party applies a sequence of instruments on the received system. 
\textbf{(b)} Network-embedding scenario: the source $\rho_{AB}$ is embedded in a larger network with two additional independent sources. 
\textbf{(c)} Broadcasting scenario: each party applies a local operation that splits the received system into two subsystems, $A_1,A_2$ and $B_1,B_2$.}
\label{fig:broadcast-vs-bowles}
\end{figure*}

An alternative line of work keeps the  devices completely uncharacterised, but modifies the \emph{causal structure} of the experiment.  This is best illustrated on the following single-copy\footnote{Yet another possibility is the many-copy activation of nonlocality \cite{Palazuelos2012,CavalcantiAcinBrunnerVertesi2013}.} examples, sketched in  Fig.~\ref{fig:broadcast-vs-bowles}.\\

\noindent\textbf{(a) Sequential measurements scenario.} 
Generalising the single–round Bell test, Alice and Bob may perform sequences of local instruments on their systems~\cite{Bowles2020boundingsetsof}. The simplest instance is hidden nonlocality~\cite{Popescu1995,Gisin1996}, where local filters are applied before a Bell test. It is now known that in the hidden nonlocality scenario there exist entangled states that remain Bell local~\cite{Hirsch2013,hirsch16}.  Whether the gap between entanglement and nonlocality persists in general sequential scenarios remains an open question.

\noindent \textbf{(b) Network-embedding  scenario.}
Another possibility is to place the source of interest $\rho_{AB}$ inside a larger network, with additional untrusted sources and parties \cite{Cavalcanti2011,Tavakoli2022}. It was shown that by embedding $\rho_{AB}$ into a four–party network with two extra independent sources, and by exploiting self–testing of Pauli measurements, one can certify the entanglement of any bipartite state in a fully device–independent way~\cite{Bowles18}. Here the gap is closed, but at the price of introducing additional sources that must be assumed independent. Jumping ahead, network-embedding has later been shown to self-test all pure multipartite entangled states~\cite{vsupic2023quantum}, a task we consider in the second part of the paper.

\noindent\textbf{(c) Broadcasting scenario.}  
A recent approach is based on broadcasting the local subsystems of the bipartite state to additional parties via fixed local operations~\cite{Bowles21}.  It was shown that this scenario reveals the  entanglement of several ``local" states~\cite{Bowles21,Boghiu23}. In particular, for the two-qubit Werner states $\rho_{AB}$, described by the visibility parameter $v$, broadcasting enables device-independent entanglement detection down to $v>0.338$~\cite{Boghiu23}, almost covering the whole range where the states are entangled $v> \nicefrac{1}{3}$. 
Interestingly, broadcast activation of a local state has recently been demonstrated experimentally in a photonic network~\cite{VillegasAguilar2024}.

\textit{Framework and assumptions.---}  A crucial modeling choice in these, and other~\cite{Lobo2024,sekatski2025}, approaches is what kind of hidden-variable description is assumed for the devices in the experiment. One possibility is to work in a purely no-signaling framework and require that parties connected by a broadcast channel share resources that are only restricted by no-signaling constraints. A different, more restrictive option is to assume that any hidden-variable model must itself be quantum, that is, based on states, channels and measurements on some (a priori unknown) Hilbert spaces. In the broadcast approach, both viewpoints lead to meaningful notions of broadcast nonlocality and to activation results, but they address slightly different questions about what kinds of classical or post-quantum explanations are being ruled out~\cite{Bowles21,Boghiu23}.
In this work, as in \cite{Bowles18}, we adopt the latter viewpoint: we assume that the devices are described by quantum theory, i.e., that the observed behaviour arises from some global quantum state, local isometries $V_X$ and local POVMs on the outputs. We show that, under this assumption, the broadcast scenario closes the gap between entanglement and Bell nonlocality. More precisely, we consider a four-party broadcast network with a single bipartite source and two branches per party. For any entangled bipartite state $\rho_{AB}$ we construct local isometries
$V_X : L(\mathcal{H}_X) \to L(\mathcal{H}_{X_1}\otimes\mathcal{H}_{X_2})$ ($X=A,B$) and local measurements for the four parties $A_1,A_2,B_1,B_2$ such that the resulting four-partite correlations cannot be reproduced in the same broadcast network with a source that is separable across the $A|B$ cut. Moreover, the same broadcast configuration yields a universal \emph{self-testing} statement: for an arbitrary (also mixed) multipartite source, the observed correlations certify, in a device-independent sense, an extracted state that reproduces the target up to a convex decomposition over locally controlled partial transpositions. 

Our construction resembles the network-embedding scheme of Ref.~\cite{Bowles18}, but it is implemented using a \emph{single}
broadcasted source rather than additional independent sources. First, each inner link $(A_1,A_2)$ and $(B_1,B_2)$ runs a Bell test~\cite{Bowles18,bowles2018}, whose maximal violation self-tests that both pairs share a maximally entangled two-qubit state, on which $A_2$ and $B_2$ perform Pauli measurements (up to local unitaries and transposition, see below). From this, we define local
\emph{extraction instruments} that map the unknown input systems to certified
qubit registers together with classical ``transposition flags'', yielding an
explicit device-independent representation of the relevant conditional states. 

Using the extracted registers, we
then (i) implement a Bell-type functional \cite{Branciard2012} whose negativity is impossible for any
separable source compatible with the same broadcast structure, thereby certifying entanglement for every entangled $\rho_{AB}$; and (ii) show that the
same statistics certify, for general multipartite sources, an extracted operator
that reconstructs the target state up to a controlled mixture of local partial
transpositions, yielding a universal notion of broadcast self-testing.

Compared to earlier broadcast-based certification schemes~\cite{Bowles21,Boghiu23}, which significantly enlarged the set of certifiable entangled states (e.g.\ for Werner and isotropic states), our result establishes that \emph{every} entangled state becomes nonlocal in a suitable quantum broadcast scenario, and further that
arbitrary multipartite quantum states admit a corresponding broadcast self-test
in the above operational sense. Conceptually, it is also closely related to the network-embedding construction of Ref.~\cite{Bowles18,vsupic2023quantum}: here the additional trusted resources are generated by self-testing within a single broadcasted bipartite source, rather than provided by extra independent sources in a larger network. We first present the argument in the qubit case, and then explain how to extend it to arbitrary finite local dimensions via standard embeddings.

\begin{figure*}[t]
    \centering
    \includegraphics[width=1\linewidth]{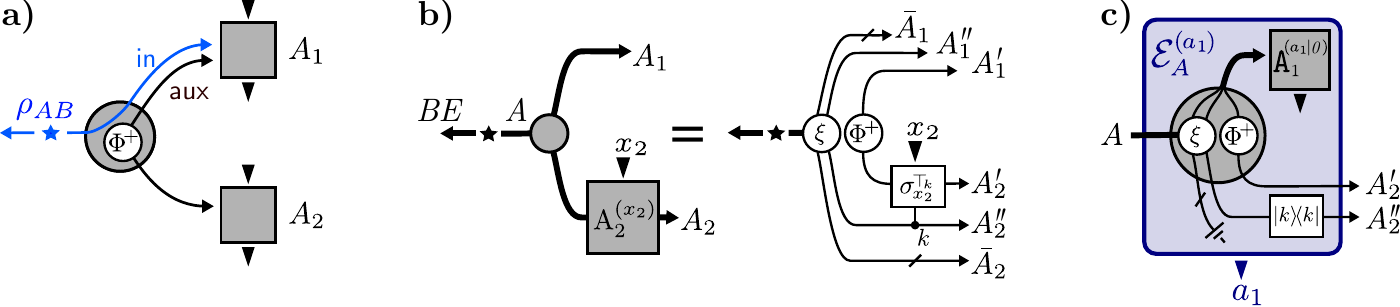}
    \caption{{\bf a)} Honest isometry in the broadcasting network enabling the device-independent 
entanglement detection of any state $\rho_{AB}$. 
Each local isometry routes the incoming system $A$ ($B$) to $A_1$ ($B_1$), 
while distributing a maximally entangled state $\Phi^+$ between 
$A_1$ ($B_1$) and $A_2$ ($B_2$). {\bf b)} Subsystem dictionary for Alice after the Bowles self-test (Eq.~\ref{eq: ST}). 
Both physical systems $A_1,A_2$  split into three subsystems.  The qubits $A_1',A_2'$ carry an EPR pair $|\Phi^+\rangle$. The remaining systems are in a state $\ket{\xi}=\sum_{k=0}^1 \ket{\xi_k}_{\bar A_1 \bar A_2 BE}\ket{kk}_{A_1''A_2''}$, where
the flag qubits $A_1'',A_2''$ label the transposition-branch, while the systems $\bar A_1, \bar A_2$ collect the remaining degrees of freedom. Up to transposition controlled by the flag qubit $A_2''$, the measurement device ${\tt A}_2^{(a_2|x_2)}$ performs the three Pauli measurements of the qubit $A_2'$. Certification of this measurement and of the EPR pair provides a
device-independent extracted qubit register $A_1'$, which serves as a certified
quantum subsystem for subsequent entanglement-detection and self-testing
arguments in the broadcast network. {\bf c)} The extraction instrument $\{\mathcal{E}_A^{(a_1)}: A \to A_2' A_2''\}$, producing two qubits and a classical outcome $a_1$ from the input system $A$.}
    \label{fig:2}
\end{figure*}

\textit{Honest broadcast experiment.---} Our broadcast map 
\begin{equation}
    V_X : X \to X_1^{\rm in} X_1^{\rm aux}X_2,
\end{equation}
follows a simple idea illustrated in Fig.~\ref{fig:2}a. Rather than ``splitting'' the incoming
system $X\in\{A,B\}$ into two subsystems, each local device  relays it to $X_1^{\rm in}$ and appends it with an EPR pair $\ket{\Phi^{+}}=\frac{1}{\sqrt{2}} (\ket{00}+\ket{11})$  distributed to $X_1^{\mathrm{aux}}$ and $X_2$.

We denote the local measurements performed by Alice and Bob respectively by the POVMs $\{{\tt A}_i^{(a_i|x_i)}\}$ and $\{{\tt B}_i^{(b_i|y_i)}\}$. In the considered scenario the measurements on the side of Alice and Bob are identical, we thus only explicitly define the former. Recall that any binary measurement ($a_i=0,1$) can be conveniently represented by
\begin{align}
    {\tt A}_i^{(a_i|x_i)}
    =: \tfrac12\bigl(\mathds{1} + (-1)^{a_i}\,{\rm A}_i^{(x_i)}\bigr)
\end{align}
where ${\rm A}_i^{(x_i)}$ is a Hermitian operator. For the honest implementation we reuse the measurement choices from the network-embedding construction of
Ref.~\cite{Bowles18}. Party $A_2$ performs the three Pauli measurements 
\begin{equation}
    {\tt A}_2^{(a_2|x_2)} = \tfrac12\bigl(\mathds{1} + (-1)^{a_2}\sigma_{x_2})=: \Pi_{A_2}^{(a_2|x_2)}, \quad x_2=1,2,3
\end{equation}
with $\sigma_{x_2}\in\{X,Y,Z\}$. Party $A_1$ carries one four-outcome ($x_1=0$) and six binary two-qubit measurements given by
\begin{align}
{\tt A}_1^{(a_1|x_1)} &=
        \Phi^{(a_1)}_{A_1^{\mathrm{aux}}A_1^{\mathrm{in}}} &\, \, & \text{for} \quad x_1=0\\
    {\rm A}_1^{(x_1)} &=  {\rm M}^{(x_1)}_{A_1^{\mathrm{aux}}}\otimes\mathds{1}_{A_1^{\mathrm{in}}}
       & \, \, & \text{for}\quad x_1 = 1,\dots,6, 
\end{align}
 with the four Bell states $\Phi^{(a_1)}\in\{\Phi^{\pm},\Psi^{\pm}\}$, and the six Hermitian operators 
\[
    {\rm M}^{(1,2)} = \frac{Z \pm X}{\sqrt{2}},\quad
    {\rm M}^{(3,4)} = \frac{Z \pm Y}{\sqrt{2}},\quad
    {\rm M}^{(5,6)} = \frac{X \pm Y}{\sqrt{2}} \,
\]

\textit{A priori model of the setup ---} 
To analyse the correlations observed in the honest implementation in a device–independent manner, we begin by
specifying the most general \emph{quantum} model compatible with the broadcasting scenario. Let $\ket{\Psi}_{ABE}$ be a purification of the unknown state $\rho_{AB}$, with an auxiliary system $E$. The splitting operations $X \to X_1X_2$ applied on both sides can always be dilated and modeled by local isometries $V_X$, allowing us to write the global state immediately before the local measurements as
\begin{equation}
    \ket{\Psi}_{A_1A_2B_1B_2E}= V_A V_B \ket{\Psi}_{ABE}.
\end{equation}
To label the measurement operators we keep the notation introduced above, with the understanding that these operators are now not assumed to be known or trusted. The correlations observed in the experiment (the behavior) are denoted $P(a_1,a_2,b_1,b_2|x_1,x_2,y_1,y_2)$, are related to the quantum model by the Born rule.\\

\textit{Self-testing of the inner links.---} In the implementation presented above, each inner link $(A_1,A_2)$ and $(B_1,B_2)$ reaches the maximal quantum score of the
Bowles test~\cite{bowles2018},
which self-tests the underlying quantum realization (Lemma 1 in~\cite{bowles2018}). As detailed in App.~\ref{app:bowles} and sketched in Fig.~\ref{fig:2}b, on Alice's side $\ket{\Psi}_{A_1A_2 BE}=V_A\ket{\Psi}_{A BE}$ this implies the following structure (up to local isometries)
\begin{align} \label{eq: ST}
&\ket{\Psi}_{A_2 A_1 BE} =  \ket{\Phi^+}_{A_2' A_1'}\ket{\xi}_{ A_2''\bar A_2 A_1''\bar A_1  BE} \\
&\tr_{A_2}[{\rm A}_2^{(x_2)}\Psi] =
\tr_{A_2}\!\left[\left(\sum_{k=0}^1(\sigma_{x_2}^{\top_{\!k}})_{A_2'}\! \otimes \ketbra{k}_{A_2''}\right) \Phi^+\!\otimes \xi\right]
 \label{eq: ST2}
\end{align}
with the auxiliary state $\ket{\xi}
=
\ket{\xi_0}_{\bar A_2 \bar A_1 BE}\ket{00}_{A_2''A_1''}
+
\ket{\xi_1}_{\bar A_2 \bar A_1 BE}\ket{11}_{A_2''A_1''}$, and ${\top_{\!k}}$ is transposing the operator if $k=1$ (${\top_{\!0}}=1,{\top_{\!1}}=\top$). Here, each system $A_i = A_i' A_i'' \bar A_i$ decomposes into three subsystems.
The qubits $A_2' A_1'$ carry an EPR pair $\Phi^+$.  
The flag qubits $A_2'' A_1''$ encode a delocalized transposition-flag $k$; which, in particular, labels the two representations $(X, \pm Y, Z)$ of the Pauli algebra related by transposition, that cannot be distinguished from the observed statistics.   Finally, the systems $\bar A_1 \bar A_2$ collect all remaining degrees of freedom. Importantly, Eq.~\eqref{eq: ST2} grants that the measurement devices ${\tt A}_2^{(a_2|x_2)}$ act on the received state as Pauli measurements of the qubit $A_2'$ -- up to transposition controlled by the flag-qubit $A_2''$, i.e. $\sum_{k=0}^1(\Pi^{(a_2|x_2)}_{A_2'})^{\top_{\!k}}\!\otimes \ketbra{k}_{A_2''}$.\\

\textit{Local extraction instruments---} With this partial characterization of Alice's devices, consider the following quantum instrument 
\begin{align}
    \cE_{A}^{(a_1)}: A &\to A_2'A_2''\\
    \cdot_A &\mapsto \tr_{A_1\bar A_2} \left[({\tt A}_1^{(a_1|0)}\otimes \id_{A_2}) \mathcal{D}_{A_2''}\circ V_A[\cdot_A]\right]\nonumber
\end{align}
defined in three following operational steps sketched in Fig.~\ref{fig:2}c. Apply the splitting map $V_A$; apply the measuremnts ${\tt A}_1^{(a_1|0)}$ on the system $A_1$; discard the system $\bar A_2$ and dephase the qubit $A_2''$ with the map $\mathcal{D}[ \mkern1mu \cdot \mkern1mu]:= \sum_{k=0}^1 \ketbra{k} \cdot \ketbra{k}$. The instrument 
extracts two qubits $A_2'A_2''$ and a classical outcome $a_1$, from the unknown input system $A$.

The same steps can be repeated to self-test Bob's side of the broadcasting network, leading to the definition of the instrument $\cE_{B}^{(b_1)}$. Applying them simultaneously on the input systems gives the following state
\begin{equation}
    \varrho_{A_2'A_2''B_2'B_2''}^{(a_1,b_2)}:=\cE_{A}^{(a_1)}\!\otimes \cE_{B}^{(b_1)}[\rho_{AB}] \!= \!\bigoplus_{k,j=0}^1\! \varrho_{A_2'B_2'}^{(k,j,a_1,b_1)}
\end{equation}
where the direct sum is taken over the dephased flag-qubits $\ketbra{k}_{A_2''}$ and $\ketbra{j}_{B_2''}$. Crucially, using the form of the measurements $\tt A_2$ and $\tt B_2$ implied by local self-testings (Eq.~\ref{eq: ST2}), we deduce the following relation between the extracted states and the observed measurement statistics (for $x_1=y_1=0$)
\begin{align} 
    &P(a_2,b_2,a_1,b_1|x_2,y_2,0,0)= \tr[\Psi ({\tt A_2}^{(a_2|x_2)}\otimes \dots {\tt B}_1^{(b_1|0)})]    \nonumber\\
 &= \sum_{k,j=0}^1 \tr \left[
   (\Pi^{(a_2 |x_2)}
_{A_1'})^{\top_{\! k}}\!\otimes\! (\Pi^{(b_2 |y_2)}
_{B_1'})^{\top_{\! j}}
\varrho_{A_2'B_2'}^{(k,j,a_1,b_1)}
    \right].
    \label{eq: KEY}
\end{align}
This is the core relation that allows us to prove the two main results of the paper.\\

\textit{Universal entanglement detection ---}
For any entangled bipartite qubit state $\rho_{AB}$ there exists an entanglement
witness~\cite{Guhne2009}, that is, a Hermitian operator $W$ satisfying
$\tr[W\rho_{AB}]<0$, and $\tr[W\sigma_{AB}]\ge 0$ for all separable
states. This operator can be expanded in any tomographically complete set, in particular in the Pauli projectors 
\begin{equation}
   W=\sum_{a,b,x,y} w_{a,b,x,y} \, \Pi_A^{(a|x)}\otimes \Pi_B^{(b|y)} \,.
   \label{eq:W-Pauli}
\end{equation}
Importantly, as the set of separable states is invariant under partial transposition, the operators $W^{\top_{\!A}},W^{\top_{\!B}}$ and $W^{\top}$ are also positive on separable state, i.e., $\tr [W^{\bullet} \rho_{AB} ]<0$ implies entanglement.

Now, consider the following quantity
\begin{align}\label{eq: Istar}
  I(P)&:=\!\! \!\! \sum_{a_2,b_2,x_2,y_2} \!\! \!\! w_{a_2,b_2,x_2,y_2} P(a_2,b_2,1,1|x_2,y_2,0,0).
\end{align}
In the honest implementation, the spitting isometries inject an EPR state $\Phi^+$, while the effects ${\tt A}^{(1|0)}_1$ and ${\tt B}_1^{(1|0)}$ are projectors onto these states, it is then not difficult to see that $I(P)= \frac{1}{4}\tr [W \rho_{AB}]<0$ (see App.~\ref{app: honest witness value} for the explicit computation). Furthermore, by virtue of Eq.~\eqref{eq: KEY} we find 
\begin{equation}\label{eq: ent witness}
    I(P) = \sum_{i,j=0}^k \tr \left[W^{\top_{\!(k,j)}} \varrho_{A_2'B_2'}^{(k,j,0,0)} \right] < 0,
\end{equation}
where $\top_{\!(k,j)}$ may transpose each system $A(B)$ depending on the value $k(j)$. The total can only be negative if at least one of the summands is. Hence, at least one of the locally-extracted states $\varrho_{A_2'B_2'}^{(k,j,0,0)}$ must be entangled, implying that the state received by A and B was also entangled.

We can now also highlight the key difference with the network-embedding scenario~\cite{Bowles18}.  In our broadcasting scenario the value $I(P)<0$ is inferred from a decomposition over the transposition branches $(k,j)$ and takes the form of a sum of contributions in which a (possibly partially transposed) operator is evaluated on branch-dependent locally extracted states. In contrast to the network-embedding scenario, here the values $(k,j)$ could have been created at the central source and be globally correlated.
Nevertheless, since each branch arises from $\rho_{AB}$ by local instruments, a violation $I(P)<0$ still implies that $\rho_{AB}$ is entangled. 

This raises the natural question whether the equivalent of the network-embedded self-testing of all states~\cite{vsupic2023quantum} is possible in the local broadcasting scenario. This is indeed the case, as we now show.\\

\textit{Universal self-testing. ---} Consider a source distributing a $N$-partite state $\rho_{\dA}$ to the parties $\dA = A_1\dots A_N$ Each party performs the same honest local splitting map of Fig.~\ref{fig:2}a) and the same honest local measurements as in the bipartite scenario. We  denote the inputs (outputs) of the lower branch of $A_i$ by $x_i\, (a_i)$ and the inputs (output) of the upper branch by $z_i (s_i)$, and collect them in vectors $\bm x, \bm a$ and $\bm z,\bm s$.  Let each party apply the local extraction instrument $\cE_{A_i}^{(s_i)}$ of Fig.~\ref{fig:2}c) on the input system, the final $2N$-qubit state reads
\begin{align}
    \varrho_{\dA'\dA''}^{(\bm s)} := \bigotimes_{i=1}^N \cE^{(s_i)}_{A_i}[\rho_{\dA}] = \!\!\!\!\!\sum_{\bm k\in\{0,1\}^N} \!\!\! \varrho_{\dA'}^{(\bm k,\bm s)}\!\otimes \ketbra{\bm k}_{\dA''}.
\end{align}
 Let all upper parties perform the measurements corresponding to the Bell state measurement in the honest case ($\bm z= \bm 0$), by virtue of the Bowles test (Eq.~\ref{eq: ST2}) the behavior satisfies
\begin{equation}
    P(\bm a,\bm s|\bm x,\bm 0) = \tr 
    \left[\left(\bigotimes_{i=1}^N \Pi_{A_i'}^{(a_i|x_i)} \right) \sum_{\bm k}(\varrho_{\dA'}^{(\bm k,\bm s)})^{\top_{\! \bm k}}\right],
\end{equation}
where we moved the partial transpose to the state as compared to Eq.~\eqref{eq: KEY}.
Since the Pauli measurements are tomographically complete, these identities give us full knowledge of the extracted operators on the right hand side. In particular, for $\bm s=\bm0$ one immediately finds $\sum_{\bm k}(\varrho_{\dA'}^{(\bm k|\bm 0)})^{\top_{\! \bm k}} = \rho_{\dA'}$. For a general $\bm s$ the same identity holds, once the extraction maps are appended with the local Pauli corrections standard in teleportation (see App.~\ref{app:bowles} for details).
Hence, after discarding the registers carrying $\bm s$ the extracted states $\varrho_{\dA'\dA''}:=\sum_{\bm s} \varrho_{\dA'\dA''}^{(\bm s)}$ is proven to be of the form 
\begin{equation}
\begin{split}\label{eq: rho ST}
&\varrho_{\dA'\dA''}= \sum_{\bm k} \varrho_{\dA'}^{(\bm k)}\otimes \ketbra{\bm k}_{\dA''}\\
&\text{such that} \quad \rho_{\dA'} = \sum_{\bm k}(\varrho_{\dA'}^{(\bm k)})^{\top_{\! \bm k}}.
\end{split}
\end{equation}
In addition, for a pure genuine multipartite entangled state $\rho_{\dA} = \psi_{\dA}$, the extracted state can be further refined to the form
\begin{equation}\label{eq: ST pure}
\varrho_{\dA' \dA''} = p\,\, \psi_{\dA'} \otimes \ketbra{\bm 0}_{\dA''} + (1-p) \, \psi_{\dA'}^{\top}\otimes \ketbra{\bm 1}_{\dA''}.
\end{equation}
The proof follows from~\cite{vsupic2023quantum} and is given in Appendix~\ref{app: proof PT} for completeness. In a nutshell, it is a consequence of partially transposed density operators having a bounded spectrum ${\rm Eig}((\varrho_{\dA'}^{(\bm k)})^{\top_{\! \bm k}})\leq   \tr\,[\varrho_{\dA'}^{(\bm k)}]$. The bound can only be saturated when $\varrho_{\dA'}^{(\bm k)}$ is a product state along the bipartition labeled by $\top_{\!\bm k}$, ensuring $\tr\,[ \psi_{\dA'} (\varrho_{\dA'}^{(\bm k)})^{\top_{\! \bm k}}]< \tr\,[\varrho_{\dA'}^{(\bm k)}]$ for $\bm k \notin\{\bm 0,\bm 1\}$.\\

In summary, the relations (\ref{eq: rho ST},\ref{eq: ST pure}) define a form of self-testing, applicable to pure as well as mixed states. Remarkably, it is tight -- in the sense that any density operator $\varrho_{\dA'\dA''}$ in Eqs.~(\ref{eq: rho ST},\ref{eq: ST pure})  can reproduce the observed measurement statistics. An example enlightening the self-test for mixed states, is that of a bipartite PPT-entangled \cite{HORODECKI1997333} state $\rho_{AB}$ (generalization of the qubit case is discussed next). In the broadcasting scenario it cannot be distinguished from any density operator $\bigoplus_{k,j=0}^1 p_{kj}\, \rho_{AB}^{\top_{\!(k,j)}}$ and is self-tested to be of this form by Eqs.~(\ref{eq: rho ST}). This
ambiguity is intrinsic, reflecting the invariance of the observed correlations
under partial transposition, and cannot be resolved without additional assumptions or resources.
\\

\textit{High-dimensional extension.---}  The construction extends to arbitrary finite local dimensions, by embedding each system $A_i$  into $n_i=\lceil\log d_{A_i}\rceil$ qubits. The local isometries would distribute that many EPR pairs per wing, which can be self-tested in parallel~\cite{Coladangelo2017}. Importantly, the parallel self-testing can be adjusted (Lemma 2 in~\cite{bowles2018}) to also certify the three Pauli measurements in each of the $n_i$ qubit-blocks of $A_i'$, up to transposition controlled by a {\it single} delocalized transposition-flag qubit encoded by $A_i''$. These self-tested qubit registers provide a tomographically complete operator basis, leading to the detection of entanglement (Eq.~\ref{eq: ent witness}) and self-testing (Eqs.~\ref{eq: rho ST},\ref{eq: ST pure}).\\

\textit{Discussion.---} 
We have shown that, under the assumption that the devices are described by quantum theory, any entangled bipartite state becomes nonlocal in a suitable broadcast Bell scenario. In this sense the usual gap between entanglement and Bell nonlocality disappears once one allows each party to apply a fixed local isometry that splits its system into two subsystems. Our proof proceeds by extracting, in a device-independent way, certified qubit subsystems on the inner links, and then using them to derive universal nonlocality and broadcast-self-testing statements.

{The main interest of the result is conceptual: it shows that, assuming the validity of quantum physics, broadcast networks are already universal for detecting entanglement device-independently  and for self-testing general multipartite states without invoking additional independent sources. A caveat concerns entanglement structures beyond mere (non)separability. 
While for pure GME targets the refinement~\eqref{eq: ST pure} collapses the ambiguity to two global transposition branches, for general mixed states the decomposition~\eqref{eq: rho ST} allows contributions from several partial-transposition patterns.  Importantly, partial transposition can change finer entanglement features across a bipartition (e.g., the Schmidt number \cite{Huber2018}, so the equivalence class singled out by~\eqref{eq: rho ST} need not preserve these features of the target state\footnote{For example, there exist a family of PPT entangled states whose Schmidt number scales linearly in the local dimension, while the Schmidt number of it's partial transpose remains bounded.}. 
Understanding which notions of multipartite entanglement (e.g. GME vs. biseparable models) can still be certified robustly within this broadcast-self-testing framework is therefore an interesting open direction.
We believe that an improvement of the certificate~\eqref{eq: rho ST} could be obtained in the multi-copy scenario, where several copies of the source are broadcasted and measured together, but it can not rule out contributions with positive partial transpose.

An important open direction is robustness. In particular,  it remains to quantify to what extent entanglement-detection capability can survive when the local devices are non-ideal, and whether useful noise thresholds can be obtained via semidefinite relaxations. It would also be interesting to identify simpler broadcast configurations (fewer settings, outcomes or parties) that still achieve universality. 

In the same vein, it would be interesting to investigate the practical implications of the local broadcasting-approach for device-independent quantum key distribution, and compare it to alternative approaches, like the routed Bell scenarios \cite{Chaturvedi2024,Lobo2024,sekatski2025} or local Bell-tests \cite{Lim2013}. As the local isometries can be performed inside Alice's and Bob's labs, the broadcasting scenario has clear potential to overcome limitations arising from transmission loss: we have shown that for any nonzero transmission the correlations observed by Alice and Bob can be guaranteed {\it long-range} quantum as long as the state $\rho_{AB}$ they receive is entangled.

\begin{acknowledgements}
We thank Victor Barizien and Nicolas Brunner for discussions. We acknowledge financial support from Swiss National Science Foundations (NCCR-SwissMAP).
\end{acknowledgements}

\bibliography{biblio.bib}

\onecolumngrid
\appendix

\section{The Bowles test and extraction instruments}
\label{app:bowles}

Here, we recall the Bowles test~\cite{bowles2018} used in the main
text: the Bell inequality and the associated self-testing statement that we
invoke.

Consider measurements on systems $A_2$ and $A_1$ with outcomes $a_2,a_1\in\{0,1\}$ and
inputs $x_2\in\{1,2,3\}$ (for $A_2$) with $x_1\in\{1,\ldots,6\}$ (for $A_1$).
For the correlators
\begin{equation}
    E_{x_2,x_1}
    :=
    \sum_{a_2,a_1=\pm1} (-1)^{a_2+a_1}\, P(a_2,a_1\,|\,x_2,x_1),
\end{equation}
consider the Bell expression
\begin{align}
    \mathcal{B}
    &=
      E_{1,1} + E_{1,2} + E_{2,1} - E_{2,2}
      \nonumber\\
    &
      +\,E_{1,3} + E_{1,4} - E_{3,3} + E_{3,4}
      \nonumber\\
    &
      +\,E_{2,5} + E_{2,6} - E_{3,5} + E_{3,6}
    \label{eq:bowles-inequality}
\end{align}
Each line is a CHSH block, so the local bound is $6$, while the quantum maximum
is $\mathcal{B}=6\sqrt{2}$.  This maximal violation is achieved, for instance,
by taking three mutually anti-commuting observables (Pauli operators) on $A_2$
and the six projective measurements on $A_1$ specified in Ref.~\cite{bowles2018}, and in the main text.\\

The maximal quantum score of the test $\mathcal{B}$ observed by the parties $A_1$ and $A_2$ by performing the binary measurements ${\tt A_i}^{(a_i|x_i)}= \frac{1}{2}(\id +(-1)^a {\rm A}_i^{(x_i)})$ on a state $\ket{\Psi}_{A_1A_2BE}$ implies the following self-testing statement (see the proof in the references).

\begin{lemma}[The Bowles self-test, Lemma 1 in~\cite{bowles2018,vsupic2023quantum}] \label{lem:bowles}
The maximal score $\mathcal{B}=6\sqrt{2}$ observed by $A_2$ and $A_1$
self-tests the following quantum model. There exist local unitaries $U_i:A_i A_i'A_i''\to \bar A_i A_i'A_i''$ such that 
\begin{align}\nonumber
(U_1\otimes U_2)\ket{\Psi}_{A_2 A_1 BE} \ket{0000}_{A_1' A_1'' A_2'A_2''}
&=
\ket{\xi}_{\bar A_1 \bar A_2 A_1''A_2'' BE}\;
\ket{\Phi^+}_{A_2' A_1'}, \qquad \text{with}
\\[2mm] \nonumber
\ket{\xi}_{\bar A_1 \bar A_2 A_1''A_2'' BE}
&=
\ket{\xi_0}_{\bar A_2 \bar A_1 BE}\ket{00}_{A_2''A_1''}
+
\ket{\xi_1}_{\bar A_2 \bar A_1 BE}\ket{11}_{A_2''A_1''},
\end{align}
\vspace{-0.8 cm}
\[\begin{aligned}
(U_{1}\otimes U_{2}) ({\rm A}_2^{(1)} \ket{\Psi})  \ket{0000}
&= 
\,\,\ket{\xi}\;
X_{A_2'}\ket{\Phi^+}_{A_2' A_1'}  &= \left(\sum_{k=0}^1 X^{\top_{\!k}}_{A_2'}\otimes \ketbra{k}_{A_2''}\right)
\ket{\xi} \ket{\Phi^+}_{A_2' A_1'}
\\[2mm]
(U_{1}\otimes U_{2})\, ({\rm A}_2^{(2)} \ket{\Psi})  \ket{0000}
&=
Z_{A_2''}\ket{\xi}\;
Y_{A_2'}\ket{\Phi^+}_{A_2' A_1'} &=  \left(\sum_{k=0}^1 Y^{\top_{\!k}}_{A_2'}\otimes \ketbra{k}_{A_2''}\right)
\ket{\xi}\ket{\Phi^+}_{A_2' A_1'}
\\[2mm]
(U_{1}\otimes U_{2})\, ({\rm A}_2^{(3)} \ket{\Psi}) \ket{0000}
&=
\,\,\ket{\xi}\;
Z_{A_2'}\ket{\Phi^+}_{A_2' A_1'}&=
 \left(\sum_{k=0}^1 Z^{\top_{\!k}}_{A_2'}\otimes \ketbra{k}_{A_2''}\right)
\ket{\xi}\ket{\Phi^+}_{A_2' A_1'}
\end{aligned}
\]
where the (unnormalised) states satisfy
$\|\ket{\xi_0}\|^2+\|\ket{\xi_1}\|^2=1$ and we used the compact notation $\top_{\!k} = \begin{cases} 1 & k=0 \\
\top & k=1
\end{cases}$.
\end{lemma}

The local unitaries $U_{i}$ are constructed from the operators ${\rm A}_i^{(x_i)}$ via a black-box circuit given in  Fig.~1 of Ref.~\cite{bowles2018}. In words, the lemma certifies that each system can be decomposed in three subsystems $A_i = A_i'A_i''\bar A_i$ such that.
\begin{itemize}
\item[(i)] The qubits $A_1'A_2'$ carry an EPR pair (as already implied by any of the maximal CHSH scores in Eq.~\eqref{eq:bowles-inequality}).
\item[(ii)] The flag qubits $A_2''A_1''$ label the``transposition branch" -- they are best understood as encoding a single delocalized logical qubit $\ket{k}_{L}= \ket{kk}_{A_2''A_1''}$. Its presence is due to the fact that transposition (or complex-conjugation) of the state and all measurement operators $ \{ \Psi_{A_1A_2BE}, {\tt A}_i^{(a_i|b_i)} \} \to \{\Psi_{A_1A_2BE}^\top, {\tt A}_i^{(a_i|b_i)\top}\}$ does not change the observed measurement statistics, preserves the validity of the model, but is in general non-trivial and cannot be absorbed in local unitaries.  Hence, from an empirical perspective we can at best reason about a weighted direct sum of the quantum models $\{ \Psi_{A_1A_2BE}, {\tt A}_i^{(a_i|b_i)}\}$ and $\{\Psi_{A_1A_2BE}^\top, {\tt A}_i^{(a_i|b_i)\top}\}$ labeled by the state $\ket{k}_L$ of the delocalized logical qubit $L$. (An interested reader may note that a similar but different equivalence has been recently discovered for the self-testing of fermionic systems via the Jordan-Wigner representation~\cite{michalski2025certifying}.) 

\item[(iii)] The systems $\bar A_1$ and $\bar A_2$ collect all remaining degrees of freedom.
\item[(iv)] The Hermitian operators $({\rm A}^{(1)}_2,{\rm A}^{(2)}_2, {\rm A}^{(3)}_2)$ act on the input state as one of the two irreps of the Pauli algebra $(X,\pm Y,Z)$ (related by transposition) applied on the qubit $A_2'$. 
The sign of the $\pm Y$ operator is precisely labeled (or controlled) by the delocalized flag qubit. The self-testing of $A_2$ measuremnts can be compactly rewritten as 
\begin{align}
(U_{A_1}\otimes U_{A_2})\, ({\rm A}_2^{(x_2)} \ket{\Psi}_{A_1A_2BE}) \ket{0000}_{A_1'A_1''A_2'A_2''}
=
 \left(\sum_{k=0}^1 (\sigma_{x_2}^{\top_{\!k}})_{A_2'}\otimes \ketbra{k}_{A_2''}\right)
\ket{\xi}_{\bar A_1 \bar A_2 A_1''A_2'' BE}\ket{\Phi^+}_{A_2' A_1'}.
\end{align}
Multiplying by $(U_{1}^\dag\otimes U_{2}^\dag)$ and using $ (U_{1}^\dag\otimes U_{2}^\dag)\ket{\xi}\ket{\Phi^+}= \ket{\Psi} \ket{0000}$, we can further rewrite this expression as 
\begin{align}
 ({\rm A}_2^{(x_2)} \ket{\Psi}_{A_1A_2BE}) \ket{0000}_{A_1'A_1''A_2'A_2''}
&=
(U_1^\dag \otimes U_2^\dag) \left(  \sum_{k=0}^1 (\sigma_{x_2}^{\top^k})_{A_2'}\otimes \ketbra{k}_{A_2''} \right)
\ket{\xi}_{\bar A_1 \bar A_2 A_1''A_2'' BE}\ket{\Phi^+}_{A_2' A_1'} \\
&=  U_2^\dag \left( \sum_{k=0}^1 (\sigma_{x_2}^{\top_{\!k}})_{A_2'}\otimes \ketbra{k}_{A_2''} \right) U_2 \ket{\Psi}_{A_1A_2BE} \ket{0000}_{A_1'A_1''A_2'A_2''}
\\
\implies \quad \tr_{A_2}[ {\tt A}_2^{(a_2|x_2)} V_A \rho_{AB} V_A^\dag ] &= \tr_{A_2} [\left( \sum_{k=0}^1 (\Pi^{(a_2|x_2)}_{A_2'})^{\top_{\!k}}\otimes \ketbra{k}_{A_2''} \right) U_2 (V_A \rho_{AB} V_A^\dag \otimes\ketbra{00}_{A_2'A_2''}) U_2^\dag \\ &=
\tr_{A_2} [\left( \sum_{k=0}^1 (\Pi^{(a_2|x_2)}_{A_2'})^{\top_{\!k}}\otimes \ketbra{k}_{A_2''} \right) \mathcal{U}_2 [V_A \rho_{AB} V_A^\dag]
].\label{eqapp: A7}
\end{align}
where we defined the isometry $\mathcal{U}_2: A_2 \to \bar A_2 A_2'A_2''$ given by $\mathcal{U}_2[\cdot_{A_2}]:= U_2(\cdot_{A_2}\otimes\ketbra{00}_{A_2'A_2''}) U_2^\dag$
\end{itemize}

With this formal definition of the self-testing, let us formally express the extraction instrument $\mathcal{E}_A^{(a_1)}:A\to A_2'A_2''$, defined operationally in the main text. Recall that it consist of three steps, apply the splitting map $V_A$, measure the system $A_1$  with ${\tt A}_1^{(a_1|0)}$, discard the system $\bar A_2$ and dephase the qubit $A_2''$ with the channel $\mathcal{D}_{A_2''}$:
\begin{align}
    \cE_A^{(a_1)}[\cdot_A] := \tr_{A_1 \bar{A}_2} \left[ ({\tt A}_1^{(a_1|0)}\otimes \id_{\bar A_2 A_2' A_2''}) \, \mathcal{D}_{A_2''}\circ \mathcal{U}_2[(V_A \cdot_A V_A^\dag)]\right].
\end{align}
Combining this with the Eq.~\eqref{eqapp: A7} gives
\begin{equation}
     \tr_{A_2'A_2''}  \left[ \left( \sum_{k=0}^1 (\Pi^{(a_2|x_2)}_{A_2'})^{\top_{\!k}}\otimes \ketbra{k}_{A_2''} ]\right)  \cE_A^{(a_1)}[\rho_{AB}]\right] = \tr_{A_2A_1}[ ({\tt A}_2^{(a_2|x_2)}\otimes {\tt A}_{1}^{(a_1|0)}) V_A \rho_{AB} V_A^\dag ].
\end{equation}
When the local Bowles test saturates its quantum maximum the same relation holds for Bob, and for any party in the multipartite setting.\\

Finally, we explicit the Pauli corrections that have to be performed after the extraction instrument, depending on the observed value $a_1$ ($s$ in the multipartite notation). In the honest scenario $ {\tt A}_{1}^{(a_1|0)}$ is the Bell state measurement with ($\sigma_0 =\id$)
\begin{equation}
{\tt A}_{1}^{(a_1|0)} = \Phi_{A_1^{\rm in} A_1^{\rm aux}}  = (\sigma_{a_1}\otimes \id)\Phi^{+}_{A_1^{\rm in} A_1^{\rm aux}}(\sigma_{a_1}\otimes \id).
\end{equation}
When performed on the input system and half of the EPR pair it realizes the teleportation map 
\begin{align}
    \tr_{A_1^{\rm in} A_1^{\rm aux}} \left[ {\tt A}_{1}^{(a_1|0)} (\cdot_{A^{\rm in}_1}\otimes \Phi^+_{A_1^{\rm aux}A_2})\right] = \frac{1}{4} (\sigma_{a_1} \cdot \sigma_{a_1})_{A_2} =: \frac{1}{4}\, \varsigma_{a_1}[\cdot]_{A_2}
\end{align}
Thus, in the honest scenario, conditionally on the value $a_1$, the state measured by $A_2$ corresponds to  $(\sigma_{a_1})_{A} \rho_{AB} (\sigma_{a_1})_{A}$. 

The same holds for all parties, implying that {\it conditionally} on the values  $\bm s =(s_1,\dots, s_N)$ observed on the upper branches for $\bm z =0$, the self-testing statement really gives  
\begin{equation}
\begin{split}
&\varrho_{\dA'\dA''}^{|\bm s)}= \sum_{\bm k} \varrho_{\dA'}^{(\bm k|\bm s)}\otimes \ketbra{\bm k}_{\dA''}\\
&\text{such that} \quad \big(\bigotimes_{i=1}^N \varsigma_{s_i}\big) [\rho_{\dA'}] = \sum_{\bm k}(\varrho_{\dA'}^{(\bm k|\bm s)})^{\top_{\! \bm k}}.
\end{split}
\end{equation}

In the last equation we can pull the self-adjoint Pauli unitary channels $\varsigma_{s_i}$ to the right-hand side of the equation and obtain 
\begin{align}
    \rho_{\dA'} =  \sum_{\bm k} \big(\bigotimes_{i=1}^N \varsigma_{s_i}\big) \left[(\varrho_{\dA'}^{(\bm k|\bm s)})^{\top_{\! \bm k}}\right] = \sum_{\bm k} \left(\big(\bigotimes_{i=1}^N \varsigma_{s_i}\big) \left[\varrho_{\dA'}^{(\bm k|\bm s)}\right] \right)^{\top_{\! \bm k}}
\end{align}
where we used $(\varsigma_s[\cdot])^\top = (\sigma_s \cdot \sigma_s)^\top = \sigma_s \cdot^{\top} \sigma_s$. Hence, the Pauli corrected states satisfies the same self-testing condition for all $\bm s$
\begin{equation}
\begin{split}
&\tilde \varrho_{\dA'\dA''}^{|\bm s)}:=  \big(\bigotimes_{i=1}^N \varsigma_{s_i}\big)[\varrho_{\dA'\dA''}^{|\bm s)}]= \sum_{\bm k} \big(\bigotimes_{i=1}^N \varsigma_{s_i}\big)[\varrho_{\dA'}^{(\bm k|\bm s)}]\otimes \ketbra{\bm k}_{\dA''}\\
&\text{such that} \quad  \rho_{\dA'} = \sum_{\bm k}(\tilde{\varrho}_{\dA'}^{(\bm k|\bm s)})^{\top_{\! \bm k}}.
\end{split}
\end{equation}
We can  now simply compose the local Pauli correction to each local extraction instrument to define the local extraction channel
\begin{align}
\mathcal{E}_{A_i}:= \sum_{s_i=0}^4 \varsigma_{s_i}\circ \mathcal{E}_{A_i}^{(s_i)}.
\end{align}
These channels deterministically extract the state $\varrho_{\dA'\dA''}$ discussed in the main text.

\section{Witness value in the honest scenario}

\label{app: honest witness value}
In this appendix, we compute the expected value of the witness in the honest scenario
\begin{align}
 I(P) := \sum_{a_2,b_2,x_2,y_2}  
 w_{a_2,b_2,x_2,y_2}\,P(a_2,b_2,a_1=1,b_1=1\,|\,x_2,y_2,x_1=0,y_1=0).
\end{align}
The computation essentially consists of using the Choi-Jamio{\l}kowski isomorphism several times while keeping track of which systems are transposed. We have
\begin{align}
 I(P) &=  \sum_{a_2,b_2,x_2,y_2}  
 w_{a_2,b_2,x_2,y_2}\,P(a_2,b_2,a_1=1,b_1=1\,|\,x_2,y_2,x_1=0,y_1=0)\\
  &= \sum_{a_2,b_2,x_2,y_2}    w_{a_2,b_2,x_2,y_2}  
 \tr [(\rho_{A_1^{\rm in}B_1^{\rm in}}\otimes \Phi^{+}_{A_1^{\rm aux}A_2}\otimes \Phi^{+}_{B_1^{\rm aux}B_2})
 ({\tt A}_2^{(a_2|x_2)}\otimes {\tt B}_2^{(b_2|y_2)} \otimes {\tt A}_1^{(1|0)}\otimes {\tt B}_1^{(1|0)})
 ]\\
 &= \sum_{a_2,b_2,x_2,y_2}    w_{a_2,b_2,x_2,y_2}  
 \tr [(\rho_{A_1^{\rm in}B_1^{\rm in}}\otimes \Phi^{+}_{A_1^{\rm aux}A_2}\otimes  \Phi^{+}_{B_1^{\rm aux}B_2})
 (\Pi_{A_2}^{(a_2|x_2)}\otimes \Pi_{B_2}^{(b_2|y_2)} \otimes \Phi^{+}_{A_1^{\rm in}A_1^{\rm aux}}\otimes \Phi^{+}_{B_1^{\rm in}B_1^{\rm aux}})
 ] \\
 &=  
 \tr [(\rho_{A_1^{\rm in}B_1^{\rm in}}\otimes \Phi^{+}_{A_1^{\rm aux}A_2}\otimes  \Phi^{+}_{B_1^{\rm aux}B_2})
 (W_{A_2 B_2} \otimes \Phi^{+}_{A_1^{\rm in}A_1^{\rm aux}}\otimes \Phi^{+}_{B_1^{\rm in}B_1^{\rm aux}})
 ]
\end{align}
with $W_{A_2B_2}:=\sum_{a_2,b_2,x_2,y_2}   w_{a_2,b_2,x_2,y_2} \Pi_{A_2}^{(a_2|x_2)}\otimes \Pi_{B_2}^{(b_2|y_2)}$. Using $\tr_{A_1^{\rm aux}}[ \Phi^{+}_{A_1^{\rm aux}A_2}\Phi^{+}_{A_1^{\rm in}A_1^{\rm aux}}] = \frac{1}{2}\sum_{n,m} \ketbra{n,m}{m,n}_{A_1^{\rm in} A_2}=\frac{1}{2} (\Phi^{+}_{A_1^{\rm in} A_2})^{\top_{A_2}} $ and $\tr_{B_1^{\rm aux}}[ \Phi^{+}_{B_1^{\rm aux}B_2}\Phi^{+}_{B_1^{\rm in}B_1^{\rm aux}}] = \frac{1}{2}\sum_{k,\ell} \ketbra{k,\ell}{\ell,k}_{B_1^{\rm in} B_2}=\frac{1}{2} (\Phi^{+}_{B_1^{\rm in} B_2})^{\top_{B_2}} $ we can further rewrite
\begin{align}
 I(P) &= 
  \frac{1}{4}\tr [(\rho_{A_1^{\rm in}B_1^{\rm in}}\otimes W_{A_2 B_2}) ( (\Phi^{+}_{A_1^{\rm in} A_2})^{\top_{A_2}}\otimes  (\Phi^{+}_{B_1^{\rm in} B_2})^{\top_{B_2}})
 ]\\
 &=\frac{1}{4}\sum_{n,m,k,\ell}
 \bra{m,k} \rho \ket{n,\ell} \bra{n,\ell} W \ket{m,k}\\
 & = \frac{1}{4}\tr [\rho_{AB} W_{AB}].
 \end{align}

\section{Broadcast-self-testing of all states} \label{app: proof PT}

In the main text we have shown that the correlations $P(\bm a,\bm s|\bm x, \bm z)$, observed in the honest broadcasting scenario with the input state $\rho_{\dA}$, prove the existence of the extraction maps $\cE_{A_i}$ producing the state 
\begin{align}
&\varrho_{\dA' \dA''} := \bigotimes_{i=1}^N \cE_{A_i}[\Psi]= \!\! \!\! \sum_{\bm k\in\{0,1\}^N}\!\! \!\! \varrho_{\dA'}^{(\bm k)} \otimes \ketbra{\bm k}_{\dA''} \qquad \\ &\text{such that}\qquad 
\rho_{\dA} = \sum_{\bm k\in\{0,1\}^N} (\varrho_{\dA'}^{(\bm k)})^{\top_{\!\bm k}}. 
\label{eq: ST cond2}
\end{align}
Assuming that the honest state is pure $\rho_\dA= \psi_{\dA}$ and genuine multipartite entangled (GME), let us now show that Eq.~\eqref{eq: ST cond2} implies $\varrho_{\dA'}^{(\bm k)}=0 $ for all $\bm k \notin\{\bm 0,\bm 1\}$. This result is proven in the Appendix D of Ref.~\cite{vsupic2023quantum}, here we retrace the derivation for completeness.\\

Consider a pure bipartite state $\Phi_{A B}$ with Schmidt decomposition $\ket{\Phi}_{A B}= \sum_{i=1}^m \lambda_i \ket{s_i}_{A}\ket{r_i}_{B}$. The spectral decomposition of the partially transposed operators reads
\begin{equation}\label{eq: spec PT}
    \Phi^{\top_{\!B}}= \sum_{i} \lambda_i^2 \ketbra{s_i,r_i^*} + \sum_{i\neq j} \lambda_i \lambda_j (\ketbra{\Psi_{ij}^+} -\ketbra{\Psi_{ij}^-}) \qquad \text{with} \qquad \ket{\Psi^{\pm}_{ij}} = \frac{1}{\sqrt 2} (\ket{s_i,r_j^*}\pm \ket{s_j,r_i^*}).
\end{equation}
To see this, verify that $\{\ket{s_i,r_i^*},\ket{\Psi^{\pm}_{ij}}\}$ are orthogonal eigenstates of $\Phi^{\top_{\!B}}$ (with the correct eigenvalues) by direct computation, and note that this set contains  $m+m(m-1)=m^2$ elements, thus forming a basis of the product Hilbert space $\mathcal{H}_A\otimes \mathcal{H}_B$ supporting $\ket{\Phi}_{A B}$.

Eq.~\eqref{eq: spec PT} implies that $\Phi^{\top_{\!B}}$ has eigenvalues in the interval $\left[-\frac{1}{2},1\right]$, and moreover can only admit an eigenvalue equal to one if $\lambda_i^2=1$, i.e., $\Phi^{\top_{\!B}} = \ketbra{s_i,r_i^*}$. By convexity, the same holds for the spectra of all partially transposed density operators 
\begin{equation}\label{eq: C4}
  -\frac{\id}{2} \preceq  \rho_{AB}^{\top_{\!B}} \preceq \id \qquad \text{with} \qquad  1\in  {\rm Eig}( \rho_{AB}^{\top_{\!B}})  \,\implies\, \rho_{AB}^{\top_{\!B}}= \ketbra{s,r^*}_{AB}.
\end{equation}

Next, we return to the self-testing condition~\eqref{eq: ST cond2}, which for a pure honest state reads
\begin{equation}
\psi_{\dA} =\!\!\! \sum_{\bm k\in\{0,1\}^N} \!\! (\varrho_{\dA'}^{(\bm k)})^{\top_{\!\bm k}} = \sum_{\bm k} p_{\bm k } (\tau_{\dA'}^{(\bm k)})^{\top_{\!\bm k}}  \qquad \text{with} \quad p_{\bm k} := \tr[\varrho_{\dA'}^{(\bm k)}],\quad \tau_{\dA'}^{(\bm k)} := \frac{\varrho_{\dA'}^{(\bm k)}}{p_{\bm k}}.
\end{equation}
As partially transposed density operators are bounded by the identity  $ (\tau_{\dA'}^{(\bm k)})^{\top_{\!\bm k}} \preceq \id$, it follows that $ \bra{\psi }(\tau_{\dA'}^{(\bm k)})^{\top_{\!\bm k}} \ket{\psi } \leq 1$ and
\begin{equation}
1 = \bra{\psi }\psi_{\dA'} \ket{\psi } = \sum_{\bm k} p_{\bm k } \bra{\psi }(\tau_{\dA'}^{(\bm k)})^{\top_{\!\bm k}} \ket{\psi }  \implies p_{\bm k } \bra{\psi }(\tau_{\dA'}^{(\bm k)})^{\top_{\!\bm k}} \ket{\psi }= p_{\bm k}.
\end{equation}
But by Eq.~\eqref{eq: C4}, the equality $1=\bra{\psi }(\tau_{\dA'}^{(\bm k)})^{\top_{\!\bm k}} \ket{\psi } \in {\rm Eig}((\tau_{\dA'}^{(\bm k)})^{\top_{\!\bm k}})$ is only possible if $(\tau_{\dA'}^{(\bm k)})^{\top_{\!\bm k}}$ is a product state across the bipartition labeled by $\bm k$. When this bipartition is nontrivial, i.e., $\bm k \notin\{\bm 0,\bm 1\}$, this is in manifest contradiction with $\ket{\psi}$ being a GME. Hence, in all these cases we must have $p_{\bm k}=0$, completing the proof.

\end{document}